\documentclass[twocolumn,prl,aps,showpacs,superscriptaddress]{revtex4-2}
\usepackage{graphicx}
\usepackage{color}

\begin{document}

\title{Self-Assembly of Magnetic Co Atoms on Stanene}

\author{Nitin Kumar}
\affiliation{Department of Physics, National Tsing Hua University, Hsinchu 300044, Taiwan}
\author{Ye-Shun Lan}
\affiliation{Department of Physics, National Tsing Hua University, Hsinchu 300044, Taiwan}
\author{Chia-Ju Chen}
\affiliation{Department of Physics, National Tsing Hua University, Hsinchu 300044, Taiwan}
\author{Yen-Hui Lin}
\affiliation{Department of Physics, National Tsing Hua University, Hsinchu 300044, Taiwan}
\author{Shih-Tang Huang}
\affiliation{Department of Physics, National Tsing Hua University, Hsinchu 300044, Taiwan}
\author{Horng-Tay Jeng}
\email{jeng@phys.nthu.edu.tw}
\affiliation{Department of Physics, National Tsing Hua University, Hsinchu 300044, Taiwan}
\affiliation{Center for Quantum Technology, National Tsing Hua University, Hsinchu 300044, Taiwan}
\affiliation{Physics Division, National Center for Theoretical Sciences, Hsinchu 300044, Taiwan}
\affiliation{Institute of Physics, Academia Sinica, Taipei 11529, Taiwan}
\author{Pin-Jui Hsu}
\email{pinjuihsu@phys.nthu.edu.tw}
\affiliation{Department of Physics, National Tsing Hua University, Hsinchu 300044, Taiwan}
\affiliation{Center for Quantum Technology, National Tsing Hua University, Hsinchu 300044, Taiwan}

\date{\today}


\vspace{2cm}
\begin{abstract}

We have investigated the magnetic Co atoms self-assembled on the ultraflat stanene on Cu(111) substrate by utilizing scanning tunneling microscopy/spectroscopy (STM/STS) in conjunction with density functional theory (DFT). By means of depositing Co onto the stanene/Cu(111) held at 80 K, Co atoms have developed into the monomer, dimer, and trimer structures containing one, two, and three Co atoms respectively. As per atomically resolved topographic images and bias-dependent apparent heights, the atomic structure models based on Sn atoms substituted by Co atoms have been deduced, which are in agreement with both self-consistent DFT calculations and STM simulations. Apart from that, the projected density of states (PDOS) has revealed a minimum at around $-0.5$~eV from the Co-3$d_{3z^{2}-r^{2}}$ minority band, which contributes predominately to the peak feature at about $-0.3$~eV in tunneling conductance ($\mathrm{d}I/\mathrm{d}U$) spectra taken at the Co atomic sites. As a result of the exchange splitting between the Co-3\textit{d} majority and minority bands, there are non-zero magnetic moments, including about 0.60 $\mu_{B}$ in monomer, 0.56 $\mu_{B}$ in dimer, and 0.29 $\mu_{B}$ in trimer of the Co atom assembly on the stanene. Such magnetic Co atom assembly therefore could provide the vital building blocks in stabilizing the local magnetism on the two-dimensional (2D) stanene with non-trivial topological properties.

\end{abstract}    

\maketitle

\section{INTRODUCTION}

Topological phases of matter have been at the center of active research since their theoretical discovery back in 1972 \cite{jm1972, jm1973, hasan2010, hasan2011, qi2011}. Topological materials provide an ideal platform for the study of fundamental physical phenomena \cite{klitzing1980, thouless1982, kane2005} and the development of technical applications such as quantum computing, \cite{kitaev2003, nayak2008} spintronics, \cite{hirohata2020} and low-power electronics.  Topological insulators (TIs) \cite{moore2010,konig2008,qi2011,fu2007PRB,qi2010,fu2007PRL,chen2011,zhang2011,kou2017} are one of the important topological quantum materials in condensed matter physics, they necessarily have topologically protected gapless states on the surface/edge in addition to the bulk bandgap \cite{hasan2010, murakami2011}. Gapless edge states provide dissipationless conducting channels as they are immune to the electron backscattering by non-magnetic defects \cite{qi2010} and disorders \cite{laughlin1981}. In the presence of the external magnetic field, the induced energy quantization of Landau-levels gives rise to the quantum Hall effect (QHE) resulting in the quantization of the Hall conductance in the units of $e^{2}/h$ (where, $e$ is the electronic charge, and $h$ is the Planck's constant) \cite{klitzing1980, thouless1982}.

Low temperature and the presence of a strong external magnetic field are required to accomplish Hall quantization in a quantum Hall insulator (QHI) \cite{li2016}, which limits their usage in practical applications. Haldane model \cite{haldane1988} proposed the spin version of the QHE referred to as the quantum spin Hall effect (QSHE) \cite{kane2005,haldane1988,kaneQSHE2005,bernevig2006,bernevig2006science,konig2007} in a two-dimensional (2D) honeycomb lattice having an alternating intrinsic magnetic field of zero net flux. Kane and Mele presented the spin-orbital coupling (SOC) in graphene as an intrinsic magnetic field \cite{kaneQSHE2005} and established graphene as the first realistic quantum spin Hall insulator (QSHI), where the QSHIs are time-reversal invariant and the corresponding topology in electronic structures is mainly driven by the combination of SOC and time-reversal symmetry \cite{hasan2010}. However, due to the weak SOC, the pristine bulk energy gap in graphene is too small (of the order of $10^{-3}$~meV) to measure directly in experiments \cite{yao2007}. On the contrary, the 2D honeycomb stanene, which is the tin (Sn) counterpart of graphene, ended up the hunt for a sizable bulk band gapped TI with an inverted gap of 300~meV \cite{xu2013, deng2018} and became an ideal choice for the study of QSHE.

Another quantum phenomenon called the quantum anomalous Hall effect (QAHE) \cite{chang2013,changMTI2013,liu2016, he2013, liu2008, deng2020, qiao2010} also has the dissipationless conducting edge feature similar to QHE without the external magnetic field and the quantization of Landau levels \cite{he2018, wang2015, nadeem2020}. Unlike the QSHE, the QAHE originates from the breaking of time-reversal symmetry as a result of the presence of an intrinsic magnetic ordering \cite{ylchen2010}. In magnetic TIs (or quantum anomalous Hall insulators), intrinsic magnetization combined with SOC drives the exchange gap accompanied by chiral edge modes which provide the dissipationless conducting channels \cite{yu2010}. Recently, several successful attempts to realize QAHE in the 3D TIs by stabilizing a long-range magnetic ordering have been reported. The magnetic bulk compound materials, such as Cr-doped (Bi$_{x}$Sb$_{1-x}$)$_{2}$Te$_{3}$ \cite{chang2013, changMTI2013}, V-doped (Bi,Sb)$_{2}$Te$_{3}$ \cite{chang2015, li2015}, MnBi$_{2}$Te$_{4}$ \cite{deng2020, otrokov2019} and, MnBi$_{2}$Te$_{7}$ \cite{vidal2019}  are a few examples where QAHE naturally occurs. Since the TIs of the Bi$_{2}$Se$_{3}$ family have a simple surface Dirac cone structure with a large bulk band gap up to 0.3~eV \cite{zhang2009, chen2009}, the adjacent ferromagnetic layers in (Bi$_{2}$Te$_{3}$)$_{n}$(MnBi$_{2}$Te$_{4}$) coupled in an antiparallel manner and exhibit the antiferromagnetic spin texture in the bulk form and demonstrate the QAHE only with an odd number of layers \cite{deng2020, liu2021, liu2020}.

Recently, in order to realize the QAHE down to the 2D limit, the ultrathin film TIs with atomic thickness have received much attention. Since the QSHE can be considered as two copies of the QAHE with magnetization vectors in opposite directions, the existence of magnetization further drives one copy to the topological trivial phase and leaves another one in the non-trivial topological phase, which transforms a non-magnetic TI into a magnetic TI \cite{chang2013,changMTI2013,deng2020,gong2019}. Based on this mechanism, introducing magnetism to the atomic-thick topological materials becomes an important issue and is worth spending effort to carry out studies down to the atomic scale.

%%%%%%%%%%%%% Here we show %%%%%%%%%%%%%%%%%%%%%%

In this presented work, we have successfully fabricated the magnetic Co atom assembly on the ultraflat stanene on Cu(111) substrate held at 80 K. We have investigated the surface structures and electronic properties by using scanning tunneling microscopy/spectroscopy (STM/STS) combined with density functional theory (DFT). In the dilute limit of coverage, the Co atoms have replaced the Sn atoms and developed into the monomer, dimer, and trimer structures on the stanene. Furthermore, the atomic structure models for magnetic Co assembly have been deduced based on the bias-dependent atomic resolution images and apparent heights, which are in line with the self-consistent lattice relaxations in DFT as well as the topographic images and line profiles in STM simulations. Besides that, the tunneling $\mathrm{d}I/\mathrm{d}U$ spectra have resolved a conductance peak at about $-0.3$~eV on the Co atomic sites, which is mainly dominated by the minimum of the PDOS at around $-0.5$~eV from the Co-3$d_{3z^{2}-r^{2}}$ minority band. Due to the exchange-split majority and minority Co-3\textit{d} bands, the net magnetic moments, i.e., about 0.60 $\mu_{B}$ in monomer, 0.56 $\mu_{B}$ in dimer, and 0.29 $\mu_{B}$ in trimer, remain presence in the Co atomic structures. Given the magnetic keystones constructed by the atomic-scale Co assembly, the local magnetism thus has an opportunity to be stabilized on the 2D topological non-trivial stanene.

%%%%%%%%%%%%%%%%%%%%%%%%%%%%%%%%%%%%%%%%%%%%%%%%%%

\section{EXPERIMENTAL AND THEORETICAL METHODS}

\textit{Experimental details.}
The whole experiment was performed in the ultra-high vacuum (UHV) environment of order $10^{-10}$ mbar. The Cu(111) substrate was cleaned by several cycles of $Ar^{+}$ sputtering and annealing. High purity Tin (Sn) (99.99\%) was evaporated from a PBN crucible in an e-beam evaporator (FOCUS) while keeping the substrate at a low temperature ($T\approx80$K). Deposition of Sn was followed by the evaporation of Cobalt (Co) atoms by heating a Co-wire (99.99\%) in an e-beam evaporator (FOCUS-EFM3). After the preparation, sample was transferred to the measurement chamber where a low-temperature scanning tunneling microscope (LT-STM) (UNISOKU, USM-1500), with a base temperature of 4.2 K, was employed to characterize the sample. All topographic images were scanned in constant current mode and \textit{in-situ} tunneling conductance ($\mathrm{d}I/\mathrm{d}U$) spectra and maps were recorded with the help of an external lock-in amplifier (Stanford Research Systems) at frequency 3991 Hz and 20 to 50 mV voltage modulation.

\textit{Theoretical computations.}
First-principles calculations were performed using the Vienna Ab Initio Simulation Package (VASP) based on Density Functional Theory (DFT). The generalized gradient approximation (GGA) in the Perdew-Burke-Ernzerhof (PBE) \cite{GKresse,GKresse1,GKresse2} form was used for exchange-correlation potential in the projector augmented wave (PAW) pseudopotential \cite{GKresse3}. The plane wave cutoff energy of 273~eV was adopted in the slab model calculations with the vacuum layer thickness of 25~{\AA} well separating the slabs. Single layer $2\times2$ supercell of stanene on top of 7-layer $4\times4$ supercell of Cu(111) was used for the Co-monomer model. While monolayer $3\times3$ stanene/7-layer $6\times6$ Cu(111) was used for Co-dimer and Co-trimer models. Then we properly replaced the Sn atoms with the Co atoms and carried out geometry optimization of the large super-structure using the gamma point for the \textit{k}-space until the total energy and the residual atomic force were converged within $10^{-4}$~eV and $-0.005$~eV/ {\AA}, respectively. The self-consistent calculations of the relaxed lattice structure were performed using $6\times6\times1$ Monkhorst-Pack \textit{k}-grid mesh with the spin-orbital coupling (SOC) included.

\section{RESULTS AND DISCUSSION}
%%%%%%%%%%%%%%%%% figure_1 %%%%%%%%%%%%%%%%%%%%%
%\begin{turnpage}
\begin{figure}[h]
\graphicspath{ {./figures/} }
\includegraphics[width=\columnwidth]{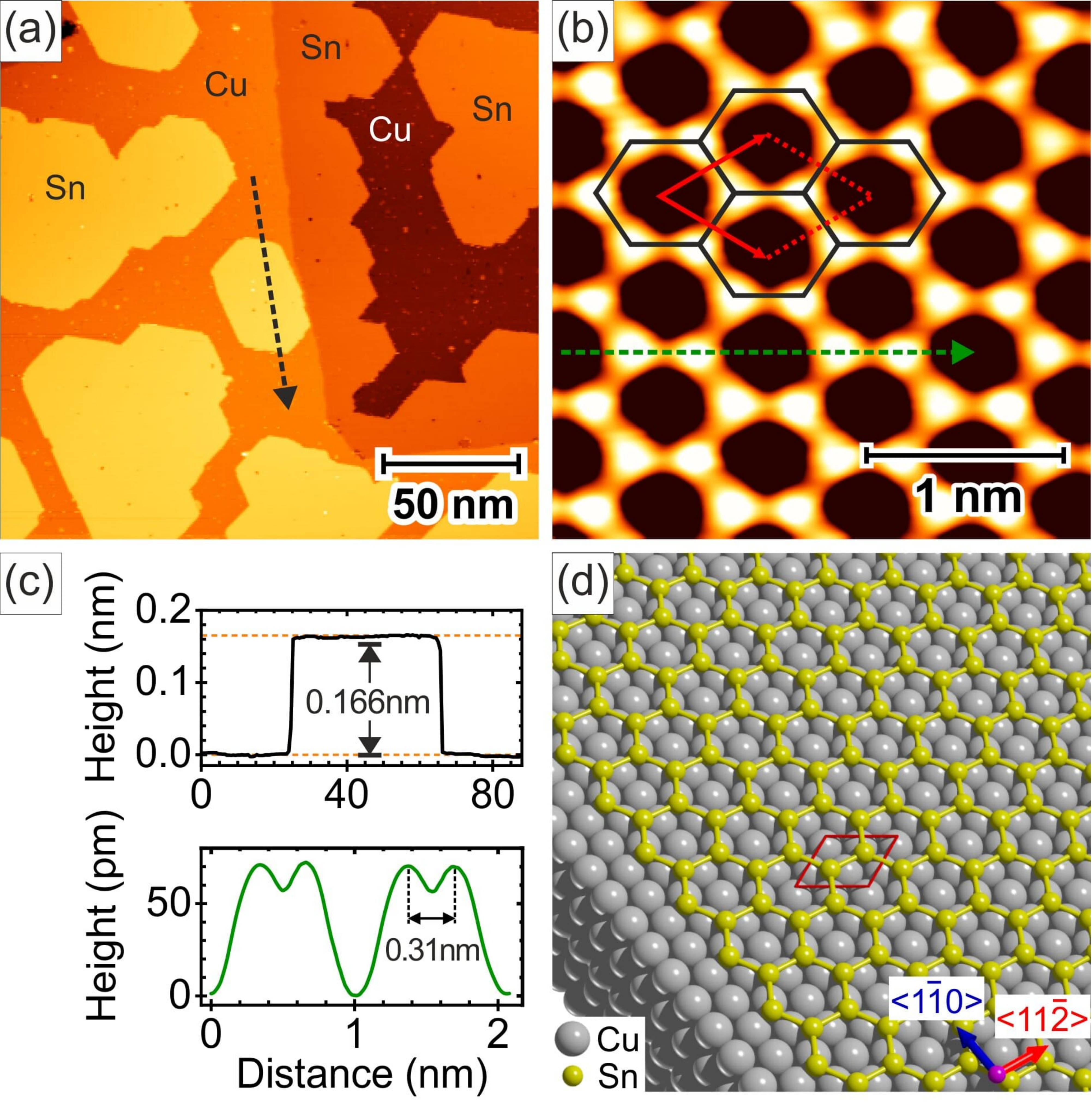}
\caption{
\label{fig1}
Ultraflat stanene on Cu(111). \textbf{(a)} An overview STM image of 2D islands of stanene on Cu(111) substrate. \textbf{(b)} Zoom-in image of one stanene island showing the flat honeycomb structure. The unit cell of stanene with the lattice constant of 0.51 nm is marked by red arrow lines. \textbf{(c)} Line profiles along the black and green dotted arrows in (a) and (b) respectively showing that the height of the stanene island is 0.16 nm and Sn-Sn bond length is 0.31 nm. \textbf{(d)} Structure model of flat honeycomb stanene on Cu(111) substrate. Grey and green balls represent the Cu and Sn atoms, respectively. (Scan parameters for (a) $V_{s} = 200$ mV, $I_{s} = 400$ pA, (b) $V_{s} = 200$ mV, $I_{s} = 1$ nA.)
}
\end{figure}
%\end{turnpage}
%%%%%%%%%%%%%%%%%%%%%% end figure_1 %%%%%%%%%%%%%%%%%%%%%%%%%%%%%

%%%%%%%%%%%%%%%%% Description of figure 1 %%%%%%%%%%%%%%%%%%%%%%

The ultraflat stanene was prepared following the analogous procedure as described in work done by Deng \textit{et al}. \cite{deng2018}, the clean Cu(111) substrate was first cooled down to 80 K on a cooling manipulator and then Sn was evaporated from a thermal evaporator. The flat 2D stanene grew in uniform and large size islands on Cu(111) was characterized by a low-temperature STM (LT-STM) at 4.2 K. Fig. \ref{fig1}(a) shows an overview of 2D stanene islands, around a two-third portion of Cu(111) substrate surface has been covered. The step height of the 2D stanene island was measured as $0.166\pm0.02$ nm  (see the top panel in Fig. \ref{fig1}(c)) from the topographic line profile along the black arrow line in Fig. \ref{fig1}(a). Six Sn atoms in the honeycomb structure can be clearly seen in Fig. \ref{fig1}(b) of atomic resolution image. The Sn-Sn bond length was measured to be 0.31 nm from the topographic line profile in Fig. \ref{fig1}(c) (bottom panel) taken along the green arrow in Fig. \ref{fig1}(b), which is close to the bond length in freestanding stanene (0.28 nm) \cite{xu2013}. The topographic line profile in Fig. \ref{fig1}(c) (bottom panel) also demonstrates that the height of all Sn atoms is equivalent, which confirms the flatness of the stanene.  The lattice constant of honeycomb stanene was measured to be 0.51 nm which is in line with the double of Cu(111) lattice constant (0.255 nm). The honeycomb unit cell is marked with red arrow lines in Fig. \ref{fig1}(b) and red rhombus in \ref{fig1}(d), which comprises two Sn atoms absorbed on $2\times2$ Cu(111) supercell and situated on hcp and fcc hollow sites.
The structure model for honeycomb stanene has been shown in Fig. \ref{fig1}(d) with a termination at the zigzag edge as reported before\cite{deng2018}.

%%%%%%%%%%%%%%%%%%%%%% figure_2 %%%%%%%%%%%%%%%%%%%%%%%%%%
%\begin{turnpage}
\begin{figure}[h]
\graphicspath{ {./figures/} }
\includegraphics[width=\columnwidth]{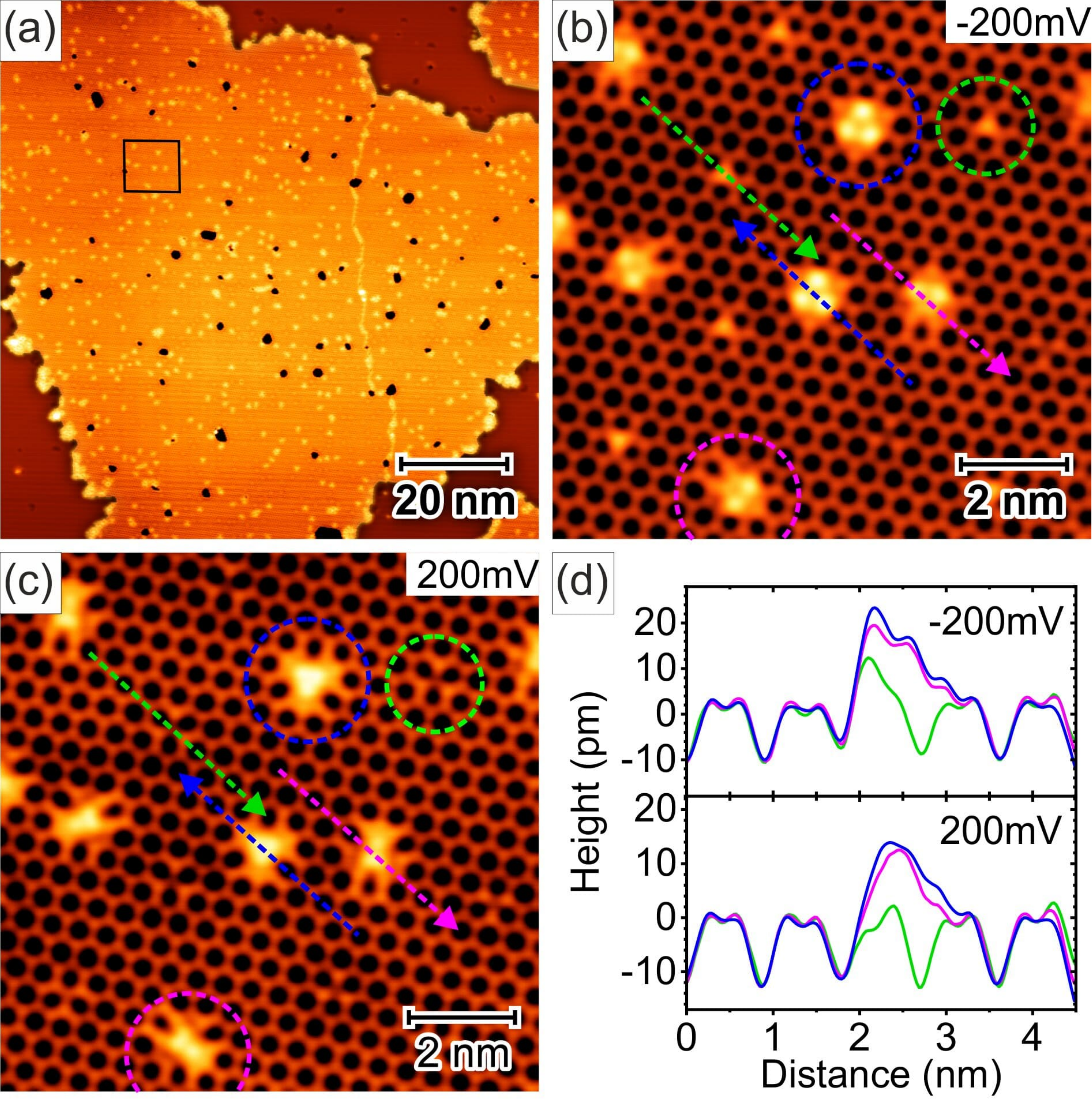}
\caption{
\label{fig2}
Self-assembled Co atomic structures on honeycomb stanene. \textbf{(a)} An overview STM image of stanene island after a cold deposition of Co atoms. \textbf{(b)} Zoom-in STM image of the black squared area in (a) taken at the negative bias voltage (filled state). The self-assembled Co atomic structures consisting of one, two, and three bright protrusions have been marked by green, magenta, and blue circles, respectively. \textbf{(c)} Zoom-in STM image of the same location with (b) taken at the positive bias voltage (empty state), where the Co atom assembly appears differently in morphology. \textbf{(d)} Bias-dependent topographic line profiles along the green, magenta, and blue arrow lines in (b) and (c). (Scan parameters for (a) $V_{s} = 1$ V, $I_{s} = 400$ pA, (b) $V_{s} = -200$ mV, $I_{s} = 1$ nA, and (c) $V_{s} = +200$ mV, $I_{s} = 1$ nA.)
}
\end{figure}
%%%%%%%%%%%%%%%%%%% end figure_2 %%%%%%%%%%%%%%%%%%%%%%%%%% 

%%%%%%%%%%%%%%%%%% Description of Fig. 2 %%%%%%%%%%%%%%%%%%

As reported before, the flat stanene is a time-reversal invariant topological material with a large SOC gap \cite{xu2013, deng2018}. In order to introduce the magnetism to the topological non-trivial stanene, we have deposited the Co atoms onto the 2D stanene islands/Cu(111). The Co atoms were evaporated at the substrate temperature at 80 K and the corresponding overview STM image has been shown in Fig. \ref{fig2}(a) where many bright protrusions, small black holes, and the irregular edge have been observed on the 2D stanene island after the Co deposition. According to the zoom-in image shown in Fig. \ref{fig2}(b) taken at the area marked by a black squared frame in Fig. \ref{fig2}(a), there are self-assembled Co atomic structures consisting of one, two, and three bright protrusions as circled by green, magenta, and blue colors, respectively. In contrast to the Fig. \ref{fig2}(b) taken at the negative bias voltage (filled state), the Fig. \ref{fig2}(c) represents the topographic image taken at the positive bias voltage (empty state) of the same scanning spot, and the Co atom assembly appears differently in morphology. The bias-dependent topographic line profiles for the Co atom assembly (green, magenta, and blue arrow lines) measured at the filled (Fig. \ref{fig2}(b)) as well as the empty (Fig. \ref{fig2}(c)) states have been arranged in Fig. \ref{fig2}(d) for a direct comparison.

%%%%%%%%%%%%%% figure_3 %%%%%%%%%%%%%%%%%%%%%%
%\begin{turnpage}
\begin{figure}[h]
\graphicspath{ {./figures/} }
\includegraphics[width=\columnwidth]{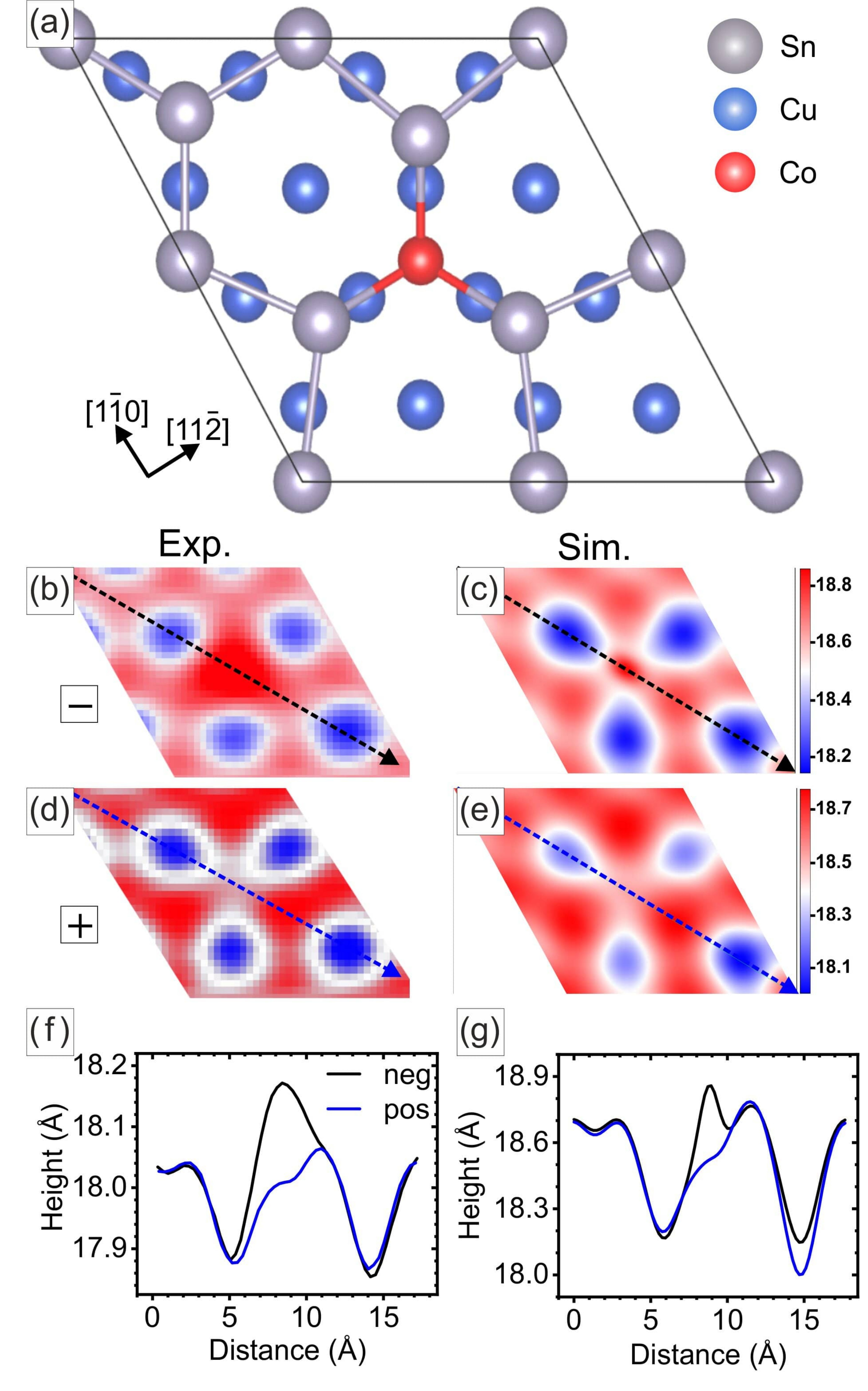}
\caption{
\label{fig3}
Co-monomer. \textbf{(a)} Proposed structure model for Co monomer from substituting one Sn atom for a Co atom in the honeycomb stanene on Cu(111) substrate. Grey, blue, and red balls in the model represent Sn, Cu, and Co atoms respectively. \textbf{(b), (d)} Atomically resolved experimental STM images of Co monomer in negative and positive sample biases. \textbf{(c), (e)} Simulated STM images based on the structure model in (a) showing an agreement with the experimental results. \textbf{(f)}, and \textbf{(g)} Bias-dependent line profiles along the black and blue dashed lines in (b)-(e).
}
\end{figure}
%%%%%%%%%%%%%%%%% end figure 3 %%%%%%%%%%%%%%%%%

%%%%%%%%%%%%%%% Description of Figure 3 %%%%%%%%%%%%%%

\textit{Monomer.}
Starting with a single bright protrusion (green circle) of three different Co atomic structures observed in Fig. \ref{fig2}(b), the energetic and hot Co atom coming from the evaporator could stand a chance to substitute one Sn atom from either hcp-hollow or fcc-hollow site of Cu(111) surface. According to the adsorption position and apparent height in the atomic resolution image of Fig. \ref{fig2}(b), the resultant proposed atomic structure model for the Co monomer by replacing one Sn atom with one Co atom in stanene has been shown in Fig. \ref{fig3}(a). The topographic appearance of Co monomer, appeared as a bright protrusion in the filled state (negative sample bias) but look like  a dip in the empty state (positive sample bias) (see \ref{fig3}(b) and (d)). The DFT calculations on self-consistent lattice relaxations show that the Co atom locates 0.51~{\AA} lower than that of stanene. However, the apparent height of the Co monomer in STM images under negative bias is around 0.1~{\AA} higher than that of stanene. This is due to the much higher density of states (DOS) of the Co-3\textit{d} band right below the Fermi level ($E_{F}$) with respect to the lower DOS of Sn as shown in Fig.~\ref{fig7}(a) and (b), giving rise to higher charge density and tunneling current for the filled states. Consequently, the apparent height of Co imaged at negative sample bias in STM topography (Fig.~\ref{fig3}(b)) is constantly higher than that of Sn, even though the position of Co is lower than Sn.
This interesting phenomenon is further supported by the STM image simulations as shown in Fig.~\ref{fig3}(c) and (e), which agree well with the experimental STM images in Fig.~\ref{fig3}(b) and (d).

To further quantify the apparent height of the Co monomer on the stanene, we have presented the experimental and theoretical line cuts (black and blue dashed lines) along the diagonal direction of the supercell under negative and positive bias voltages as shown in Fig.~\ref{fig3}(f) and (g), respectively. Five features can be found through this line, which correspond individually to the heights from Sn, Sn, Co, Sn, and Sn atoms (left to right), where they behave differently in heights for different bias voltages. It can be seen that under negative bias voltage, the Co peak is significantly higher than the Sn peak, whereas the order reverses under the positive bias voltage. Both the experimental and the theoretical height profiles agree fairly with each other.

%%%%%%%%%%%%%%% figure_4 %%%%%%%%%%%%%%%%%%%
%\begin{turnpage}
\begin{figure}[h]
\graphicspath{ {./figures/} }
\includegraphics[width=\columnwidth]{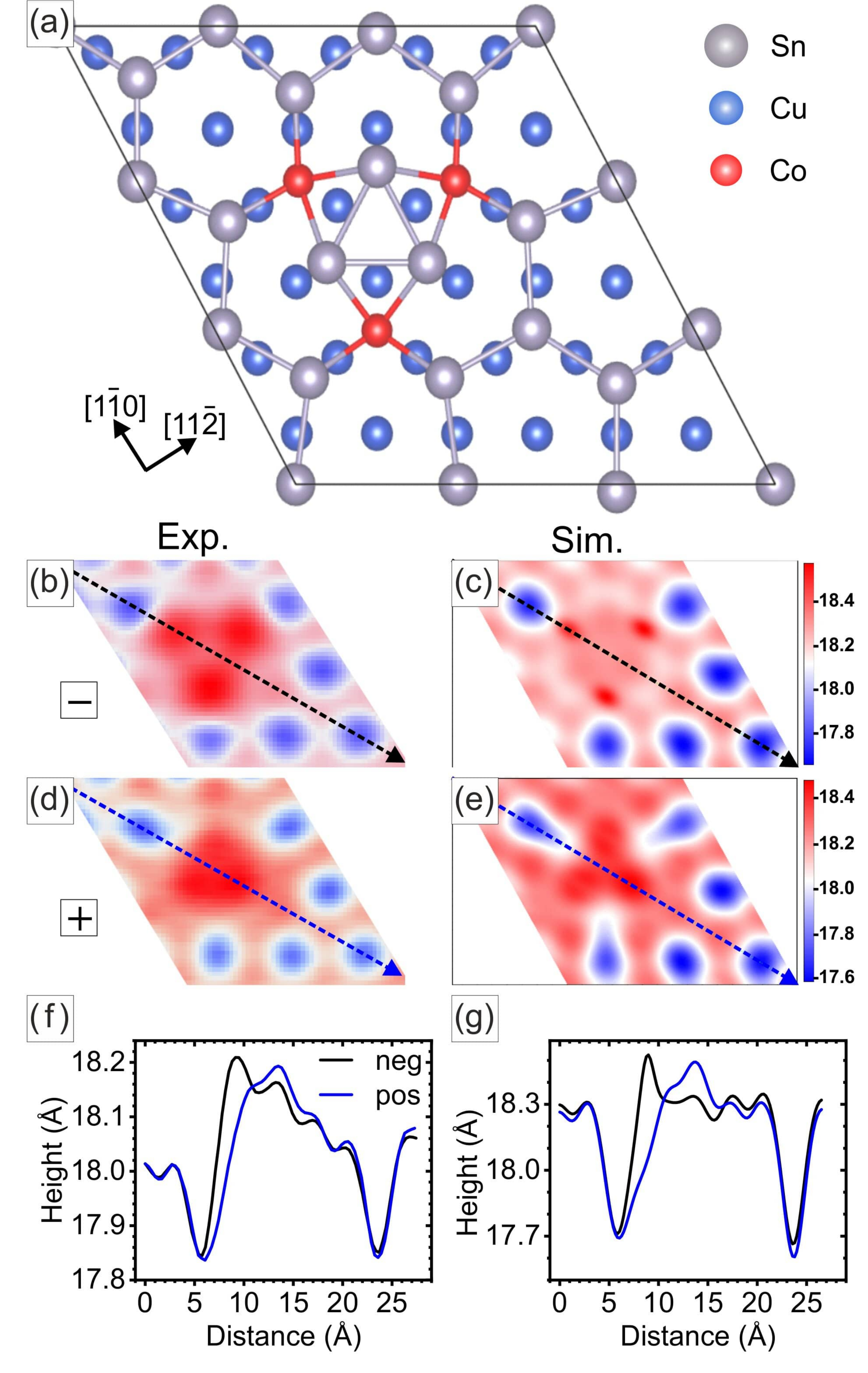}
\caption{
\label{fig5}
Co-trimer. \textbf{(a)} Proposed structure model for Co trimer. Three Co atoms have substituted three Sn atoms in the stanene and form an upside-down triangle. Grey, blue, and red balls represent Sn, Cu, and Co atoms respectively. \textbf{(b), (d)} Atomically resolved experimental STM images of Co trimer in negative and positive sample biases. \textbf{(c), (e)} Simulated STM images based on the structure model in (a) are consistent with the topographic features and symmetry observed in the experimental STM images. \textbf{(f)}, and \textbf{(g)} Bias-dependent height profiles along the black and blue dashed lines in (b)-(e).
}
\end{figure}
%%%%%%%%%%%%%%%% end figure 4 %%%%%%%%%%%%%%%

%%%%%%%%%%%%% Description of figure 4 %%%%%%%%%%%%%

\textit{Trimer.}
After the Co monomer, it's easier to discuss the high-symmetric Co atomic structure with 3 bright protrusions (blue circle) than the low-symmetric Co atomic structure with 2 bright protrusions (magenta circle) as shown in Fig.~\ref{fig2}(b). The three-fold symmetric Co atomic structure model has been constructed in two steps: First, four adjacent Sn atoms were replaced by four Co atoms, forming a big Co triangle. Second, the central Co atom was replaced by a small Sn triangle as shown in Fig.~\ref{fig5}(a), and we assign this structure as the Co trimer in the following. Similar to the monomer case, not only the structural relaxations of Co trimer have been carried out, but also the PDOS for Co trimer has been calculated as shown in Fig.~\ref{fig7}(d), showing that the Co-3\textit{d} band dominates below $E_{F}$ (filled states). The resultant bias-dependent simulated STM images have been further shown in Fig.~\ref{fig5}(c) and (e). The simulated topography displays a Co-derived upside-down triangle (Fig.~\ref{fig5}(c)) in the negative bias voltage, on the other hand, it becomes an Sn-derived triangle (Fig.~\ref{fig5}(e)) in the positive bias voltage. It's noted that not only the simulated STM images are in good agreement with the experimental STM images (Fig.~\ref{fig5}(b) and (d)), but also the corresponding height profiles are reasonably consistent with each other (Fig.~\ref{fig5}(f) and (g)).

%%%%%%%%%%%%%%%%% figure_5 %%%%%%%%%%%%%%%%%

%\begin{turnpage}
\begin{figure}[h]
\graphicspath{ {./figures/} }
\includegraphics[width=\columnwidth]{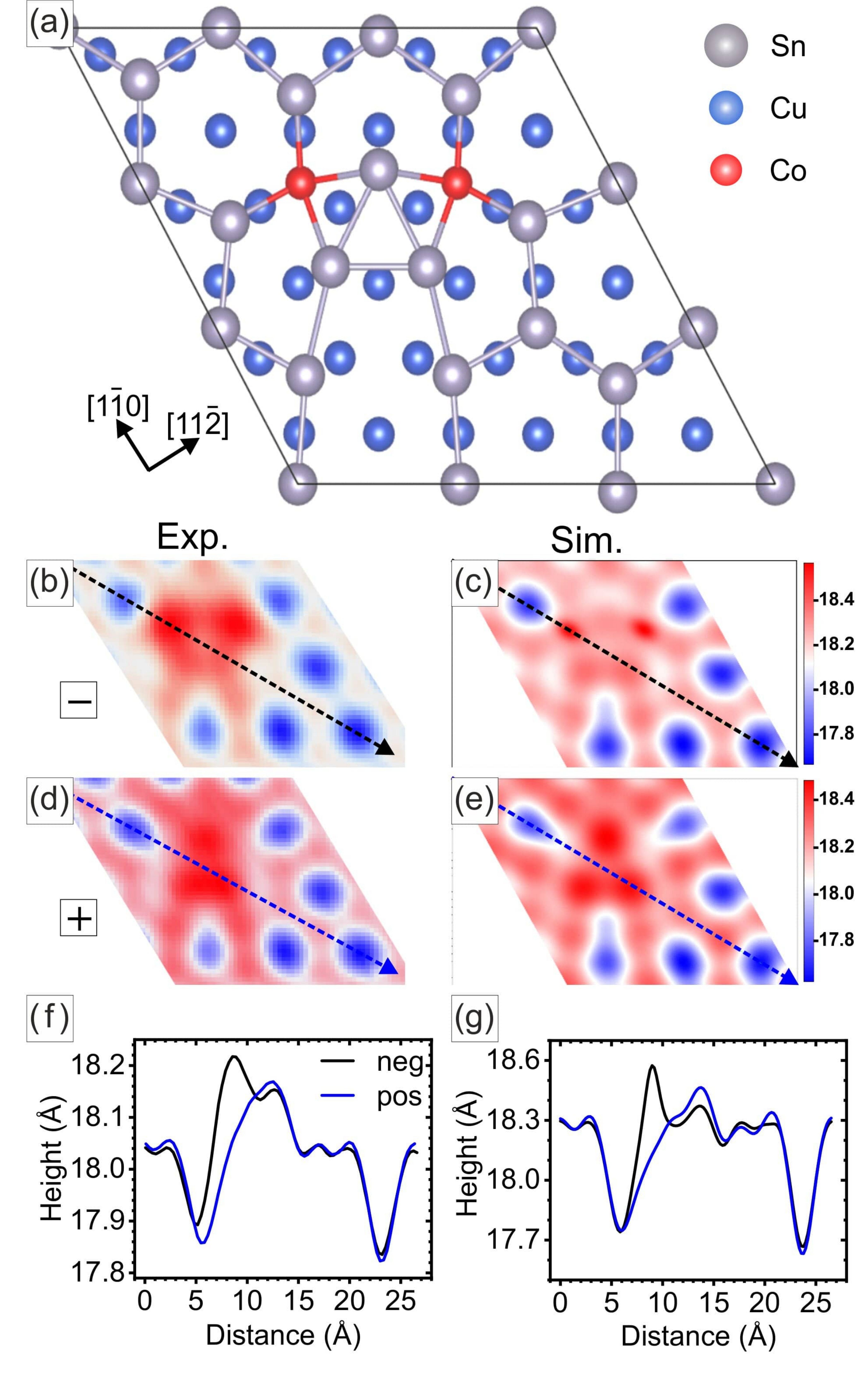}
\caption{
\label{fig4}
Co-dimer. \textbf{(a)} Proposed structure model for Co dimer. The low-symmetric Co dimer structure has been deduced from modifying the high-symmetric Co trimer structure by removing one substituted Co atom. Grey, blue, and red balls in the model represent Sn, Cu, and Co atoms respectively. \textbf{(b), (d)} Atomically resolved experimental STM images of Co dimer in negative and positive sample biases. \textbf{(c), (e)} Simulated STM images based on the structure model in (a) representing a good consistency with experimental STM images. \textbf{(f)}, and \textbf{(g)} Bias-dependent height profiles along the black and blue dashed lines in (b)-(e).
}
\end{figure}

%%%%%%%%%%%%%%%% end figure 5 %%%%%%%%%%%%%%%%%%

%%%%%%%%%%% Description of figure 5 %%%%%%%%%%%%

\textit{Dimer.}
The structure model of the low-symmetric Co atomic structure with two bright protrusions (magenta circle) in Fig.~\ref{fig2}(b) has been deduced from the high-symmetric Co trimer structure model in Fig.~\ref{fig5}(a).  By removing one Co atom from the Co trimer structure in Fig.~\ref{fig5}(a), the corresponding Co dimer structure has been shown in Fig.~\ref{fig4}(a). In this model, two Co atoms substitute two Sn atoms from the two identical positions in the honeycomb lattice. The PDOS of Co dimer has been calculated in Fig.~\ref{fig7}(c) and also shows the prominent contribution of the Co-3\textit{d} band in the energy range below $E_{F}$. The simulated STM images of the Co dimer structure under different biases show clearly different characters: In negative bias, a dumbbell-like shape contour emerges from the Co dimer region in Fig.~\ref{fig4}(c), whereas a triangle-like contour driven from the central Sn triangle appears in Fig.~\ref{fig4}(e) of the positive bias. Such bias-dependent simulated STM images support the experimental observations that a dumbbell-like Co dimer is relatively brighter in the negative bias of Fig.~\ref{fig4}(b), but the Sn triangle becomes more pronounced in the positive bias of Fig.~\ref{fig4}(d). It's also noted that the bias-dependent height profiles have been further examined and they are also consistent in both experiments and simulations as shown in  Fig.~\ref{fig4}(f) and (g).

%%%%%%%%%%%% figure_6 %%%%%%%%%%%%%%%%%%%%%%
%\begin{turnpage}
\begin{figure}[h]
\graphicspath{ {./figures/} }
\includegraphics[width=\columnwidth]{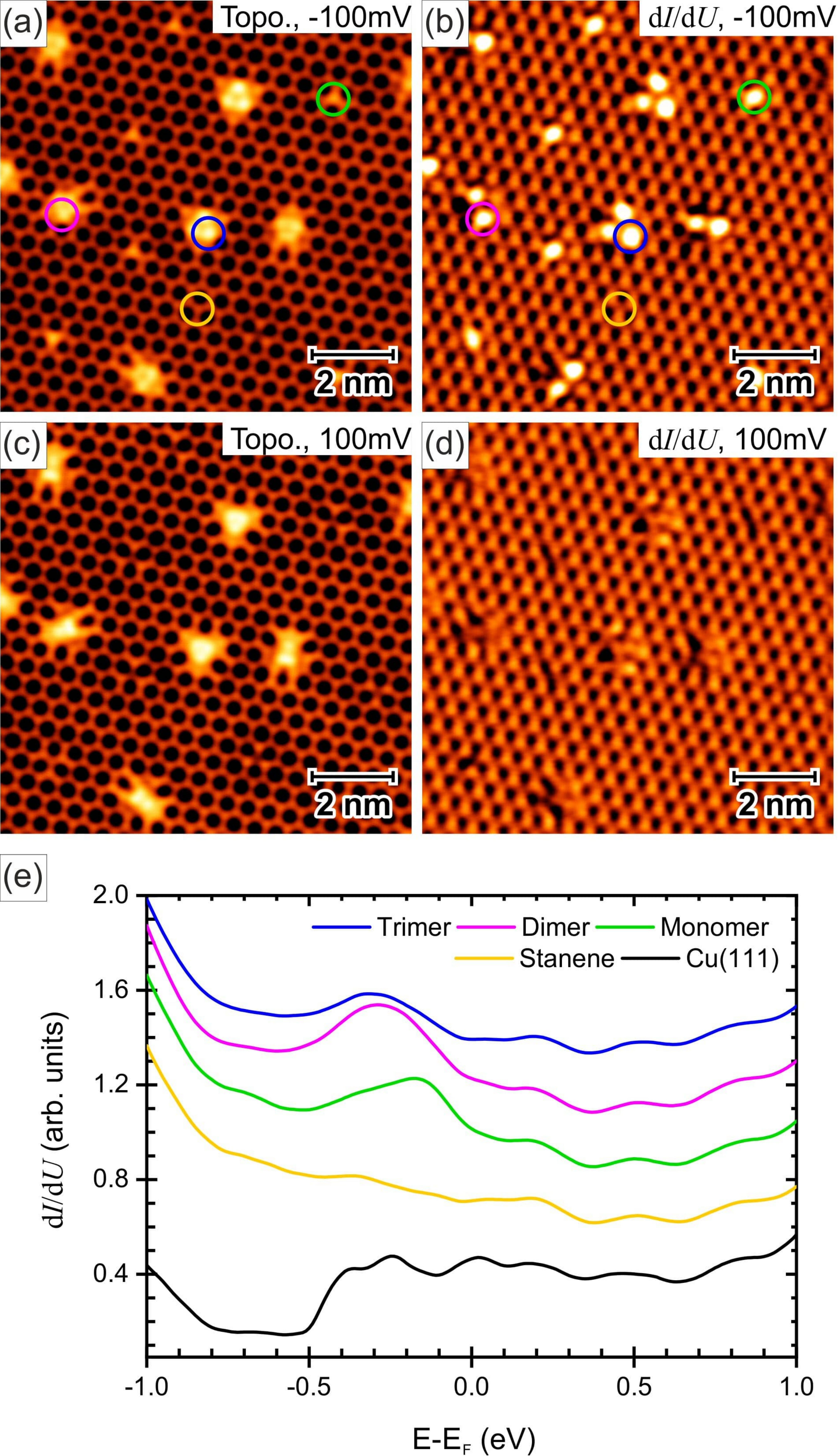}
\caption{
\label{fig6}
Bias-dependent $\mathrm{d}I/\mathrm{d}U$ maps. \textbf{(a), (c)} Filled and empty state STM topographic images of Co atom assembly on honeycomb stanene. \textbf{(b), (d)} $\mathrm{d}I/\mathrm{d}U$ maps corresponding to (a) and (c). \textbf{(e)} $\mathrm{d}I/\mathrm{d}U$ curves as a function of sample bias (positions where the $\mathrm{d}I/\mathrm{d}U$ curves are taken have been marked with corresponding colored circles in (a) and (b)). (Scan parameters for (a), (c) $V_{s} = -100, +100$ mV, $I_{s} = 1$ nA; (b), (d) $V_{s} = -100, +100$ mV, $I_{s} = 1$ nA; stabilization parameters for (e) $V_{s} = +1.0$ V, $I_{s} = 1$ nA.)
}
\end{figure}
%%%%%%%%%%%%%%%%% end figure 6 %%%%%%%%%%%%%%

%%%%%%%%%%%%%% Description of Fig. 6 %%%%%%%%%%%%%%

Fig. \ref{fig6}(a) and (c) represent the bias-dependent topographic images taken at the negative (filled state) and the positive biases (empty state), respectively. The corresponding $\mathrm{d}I/\mathrm{d}U$ maps of Co atom assembly on the stanene have been shown in Fig. \ref{fig6}(b) and (d), displaying a bias dependence as well. In order to understand the local density of states (LDOS) of these self-assembled Co atomic structures, we have exploited the STS measurements with high spatial and energy resolution. The systematic studies of $\mathrm{d}I/\mathrm{d}U$ curves acquired at the different positions have been arranged into the Fig. \ref{fig6}(e). The black line is the $\mathrm{d}I/\mathrm{d}U$ curve taken from the bare Cu(111), where the well known Shockley surface state at $-0.44$~eV has been resolved. A nearly featureless $\mathrm{d}I/\mathrm{d}U$ curve (yellow line) indicates the constant DOS of stanene, which is in line with the PDOS calculated in Fig. \ref{fig7}(a). The $\mathrm{d}I/\mathrm{d}U$ spectra measured at the Co atomic sites of Co monomer (green circle), dimer (magenta circle), and trimer (blue circle) exhibit a broad peak feature at around $-0.3$~eV. This peak feature resolved in the $\mathrm{d}I/\mathrm{d}U$ spectra also reflects in the bias-dependent $\mathrm{d}I/\mathrm{d}U$ maps as shown in Fig. \ref{fig6}(b) and \ref{fig6}(d), where Co monomer, dimer, and trimer have high conductance intensity and appear bright in the negative bias (filled state), but have no significant contrast as compared to honeycomb stanene in the positive bias (empty state).

%%%%%%%%%%%%%%% figure_7 %%%%%%%%%%%%%%%%%%%%%%
%\begin{turnpage}
\begin{figure}[h]
\graphicspath{ {./figures/} }
\includegraphics[width=\columnwidth]{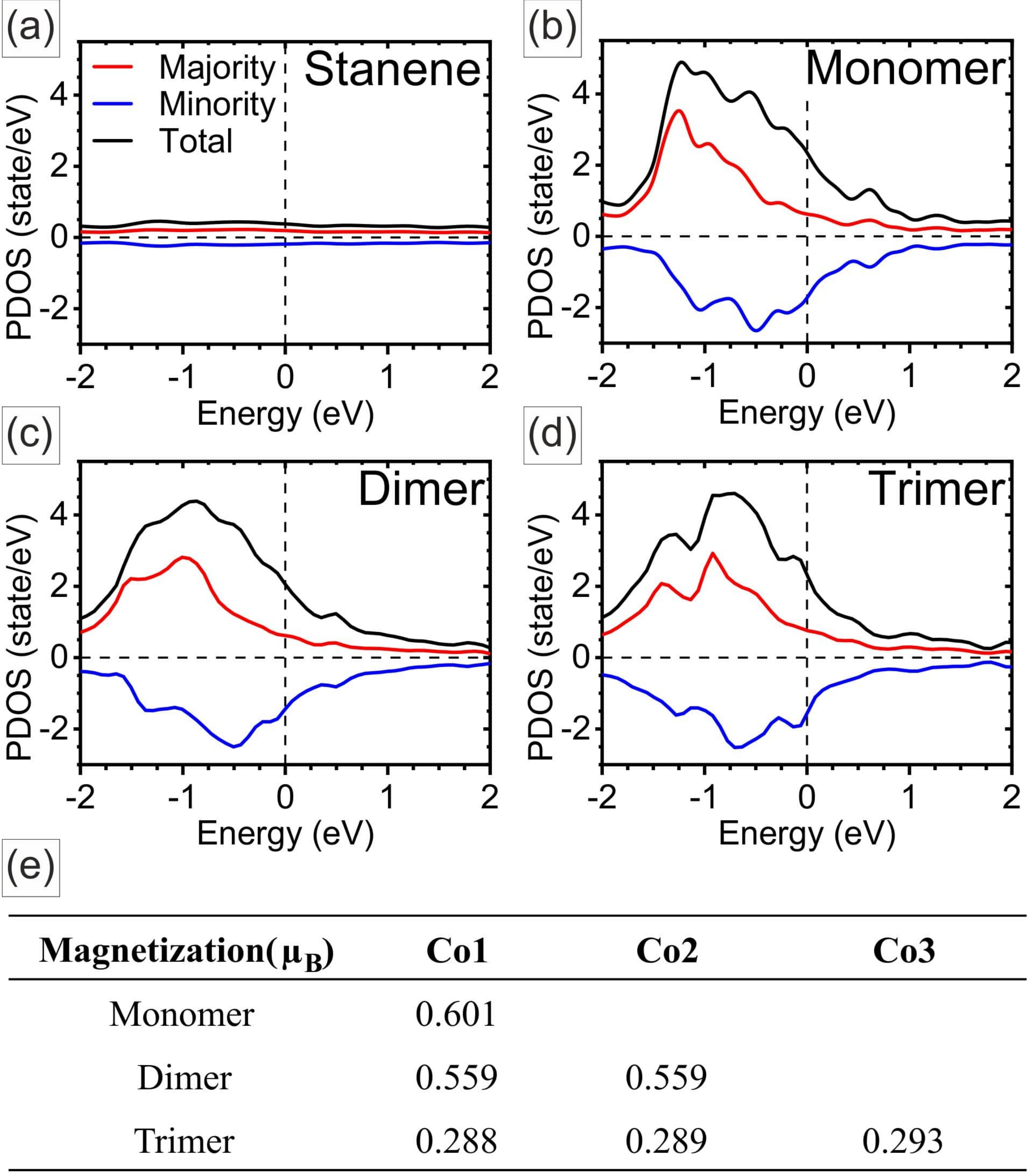}
\caption{
\label{fig7}
Projected density of states (PDOS) and magnetic moments. \textbf{(a)-(d)} PDOS curves for pristine stanene, Co monomer, dimer, and trimer. \textbf{(e)} Table of calculated magnetic moments of the individual Co atoms in monomer, dimer, and trimer structures.
}
\end{figure}
%%%%%%%%%%%%%%%%%% end figure 7 %%%%%%%%%%%%%%%%%%

%%%%%%%%%%%%%% Description of figure 7 %%%%%%%%%%%%%%

The PDOS of stanene, Co-monomer, Co-dimer, and Co-trimer have been presented in Fig.~\ref{fig7}. The intensity of Sn PDOS is relatively low and featureless in the energy dispersion of $\pm~2.0$~eV with respect to $E_{F}$. In addition, the PDOS for the Co-monomer, Co-dimer, and Co-trimer have been shown in Fig.~\ref{fig7}(b), (c), and (d), respectively, their PDOS clearly show an exchange splitting that originates from the Co-3\textit{d} majority (red) and minority (blue) bands. By taking the Co monomer for an example, the splitting can be easily identified from a maximum at $-1.25$~eV of the Co-3\textit{d} majority spin band (red line) and a minimum at around $-0.5$~eV of the Co-3\textit{d} minority spin band (blue line), and the same analogy can be applied for the cases of Co-dimer and trimer. Interestingly, we have found such exchange splitting between the Co-3\textit{d} majority and minority bands continuously reduces from the Co-monomer to the Co-trimer, implying the delocalization of \textit{d} electrons due to a greater hybridization and orbital overlapping with Sn atoms in the Co dimer and timer configurations. Given the exchange-split Co-3\textit{d} majority and minority bands, the net magnetic moments of the Co atom assembly on the stanene have been evaluated and they are nearly 0.60~$\mu_B$, 0.56~$\mu_B$, and 0.29~$\mu_B$ for the Co monomer, dimer, trimer, respectively, as summarized in the Fig.~\ref{fig7}(e). Apart from that, as the STS measurements arranged in the Fig. \ref{fig6}(e), the Co atom assembly exhibits a peak feature at around $-0.3$~eV in the $\mathrm{d}I/\mathrm{d}U$ spectra, and we can further identify this peak coming from the minimum at $-0.5$~eV of the Co-3\textit{d} minority band, which is mainly contributed from the 3$d_{3z^{2}-r^{2}}$ orbital. We denote that this conductance peak at $-0.3$~eV has been reported on Co/Cu(111)\cite{OPietzsch}, Co/Pt(111)\cite{FMeier}, Co/Ir(111)\cite{JEBickel} and Co/W(110)\cite{JWiebe} in the previous STS measurements. Despite a small deviation in energy position, this peak feature has been typically observed for the 2D Co nanoislands grown on a wide range of substrates. In combination of photoemission and DFT studies\cite{FMeier,JWiebe,LDiekhoner,FJHimpsel,HKnoppe,SNOkuno}, the \textit{d}-like minority character has been identified for this filled-state conductance peak.

%%%%%%%%%%%%%%%% DISCUSSION %%%%%%%%%%%%%%%%
As reported by Deng \textit{et al}.\cite{deng2018}, the ultraflat stanene on Cu(111) substrate have been characterized by the SOC induced topological gap along with the inverted band structure and the helical edge states at the boundary. In addition, the bulk magnetic TIs arose from the doping of magnetic atoms have been fabricated and exhibit zero-field QAHE\cite{deng2020,gong2019,changMTI2013}. In the present work, by depositing the Co atoms onto the stanene/Cu(111) at 80 K, the self-assembled Co atomic structures, e.g., monomer, dimer, and trimer, have been fabricated. Given the high spatial and energy resolution in the STS measurements, a characteristic peak feature at about $-0.3$~eV has been resolved in $\mathrm{d}I/\mathrm{d}U$ spectra at the Co atomic sites, which can be deduced from a minimum at around $-0.5$~eV of the PDOS from the Co-3$d_{3z^{2}-r^{2}}$ minority band. As a result of the exchange-split Co-3\textit{d} majority and minority bands, there are net magnetic moments of about 0.60 $\mu_{B}$ in monomer, 0.56 $\mu_{B}$ in dimer, and 0.29 $\mu_{B}$ in trimer on the stanene/Cu(111). In the perspective of the doping stanene with Co impurities from our studies, such magnetic Co assembly has provided a good starting point to develop atomic-scale magnetism in the 2D stanene/Cu(111) with non-trivial topological properties.

\section{SUMMARY}

In summary, the self-assembled Co atomic structures have been fabricated on the ultraflat stanene/Cu(111) by means of the low temperature growth. On account of atomically resolved topographic images and bias-dependent apparent heights, the Co monomer, dimer, and trimer structures from substituting Sn atoms have been deduced, which are in agreement with the self-consistent structural relaxations in DFT and have been further supported by the STM simulations. In addition, the STS measurements with high spatial and energy resolution have resolved a conductance peak at about $-0.3$~eV in the $\mathrm{d}I/\mathrm{d}U$ spectra acquired at the Co atomic sites, which can be explained by a minimum at around $-0.5$~eV of the PDOS from the Co-3$d_{3z^{2}-r^{2}}$ minority band. Since there is an exchange splitting between the Co-3\textit{d} majority and minority bands, non-zero magnetic moments have been found in the Co atom assembly, ranging from about 0.60 $\mu_{B}$ in monomer, to about 0.56 $\mu_{B}$ in dimer, and to about 0.29 $\mu_{B}$ in trimer on the stanene. These Co atomic structures, therefore, serve as the atomic-scale magnetic dopants, offering the quintessential spin ingredients to develop the local magnetism in the 2D topological non-trivial stanene.

\section{Acknowledgments}

N.K. and Y.S.L. contribute equally to this work. P.J.H. acknowledges support from the competitive research funding from National Tsing Hua University, Ministry of Science and Technology of Taiwan under Grants No. MOST-110-2636-M-007-006 and MOST-110-2124-M-A49-008-MY3, and center for quantum technology from the featured areas research center program within the framework of the higher education sprout project by the Ministry of Education (MOE) in Taiwan. J.H.T. acknowledges the financial support from the Ministry of Science and Technology of Taiwan under Grants No. MOST-109-2112-M-007 -034 -MY3 and also from the NCHC, CINC-NTU, and AS-iMATE-109-13, Taiwan.

%\bibliographystyle{apsrev4-2}
%\bibliography{references.bib}
%\end{document}

%%%%%%%%%%%%%%%%%%%%%%%%%%%%%%%%%%%%%%%%%%%%%%%%%%%%%%
%references begin
%%%%%%%%%%%%%%%%%%%%%%%%%%%%%%%%%%%%%%%%%%%%%%%%%%%%%%

%%%%%%%%%%%%%%%%%%%%%%%%%%%%%%%%%%%%%%%%%%%%%%%%%%%%%%
%references end
%%%%%%%%%%%%%%%%%%%%%%%%%%%%%%%%%%%%%%%%%%%%%%%%%%%%%%

\end{document}